# TOPIC EXTRACTION OF CRAWLED DOCUMENTS COLLECTION USING CORRELATED TOPIC MODEL IN MAPREDUCE FRAMEWORK

Mi Khine Oo[1] and May Aye Khine[2]

[1]Numerical Analysis Lab, University of Computer Studies, Yangon, Myanmar
[2]Faculty of Computing, University of Computer Studies, Yangon, Myanmar

## ABSTRACT

*The tremendous increase in the amount of available research documents impels researchers to propose topic models to extract the latent semantic themes of a documents collection. However, how to extract the hidden topics of the documents collection has become a crucial task for many topic model applications. Moreover, conventional topic modeling approaches suffer from the scalability problem when the size of documents collection increases. In this paper, the Correlated Topic Model with variational Expectation-Maximization algorithm is implemented in MapReduce framework to solve the scalability problem. The proposed approach utilizes the dataset crawled from the public digital library. In addition, the full-texts of the crawled documents are analysed to enhance the accuracy of MapReduce CTM. The experiments are conducted to demonstrate the performance of the proposed algorithm. From the evaluation, the proposed approach has a comparable performance in terms of topic coherences with LDA implemented in MapReduce framework.*

## 1. INTRODUCTION

With increased online digital documents, researchers have started to focus on large documents collection for the extraction of hidden semantic themes and the summarization of these large collection. As more and more digitized documents are spreading and scattering across many sources, such as blogs and websites, it has become important to gather these documents and examine valuable data from these gathered documents to uncover the hidden themes.

Probabilistic topic models discover the underlying thematic structures in a collection of documents by extracting the topics. With these extracted topics, the whole documents collection can be summarized and categorized without human annotation effort. Latent Dirichlet Allocation (LDA) [2], one of the most widely known topic models, uses statistical methods to infer the latent topics contained in the document collection. A main shortcoming of LDA is the lack of ability to model the correlations between topics because of using a Dirichlet distribution in order to model the topic proportions. The distribution assumes that the existence of one topic is not correlated with the existence of another because of its independence structure. However, the latent topics can have correlations between each other in many practical applications. Hence, Correlated Topic Model (CTM) [3] proposed a solution to solve the incapability of LDA by substituting the Dirichlet distribution with the logistic normal distribution to exhibit the correlations of the latent topics.

The CTM may have a challenge in calculating the posterior distribution of topics over the observed words when inferring the latent topics. To figure out the model parameters estimation for a topic model, different inference algorithms including Gibbs Sampling and Variational Expectation-Maximization (EM) have been introduced [6]. Gibbs sampling is a Markov Chain Monte Carlo algorithm which draws samples from probability distributions. Variational EM algorithm relies in computing the maximum likelihood estimates of parameters [3]. In this work, the variational EM algorithm is applied for the analysis.

The previous studies of LDA utilized distributed computational resources with different parallelized algorithms. Nallapati et al. [14] proposed a parallelized variational EM algorithm for LDA in multiprocessor and distributed implementations. Wolfe et al. [7] presented a fully distributed EM framework to distribute the computation and parameter storage across three Network topologies. Moreover, Newman et al. [1] proposed two distributed inference algorithms using Gibbs Sampling technique for LDA to distribute the data and parameters over distinct processors.

When the advent of large-scale processing platforms comes out, the studies of LDA in MapReduce framework are introduced in a number of works. The authors of [9] used the variational inference technique to propose a parallelized Mr. LDA algorithm and implemented the algorithm in MapReduce framework. In [13], the author proposed a novel MapReduce based framework by utilizing K-means clustering and LDA topic model to summarize the large text collection. Furthermore, reference [17] proposed a novel model Mr. sLDA which extends the supervised LDA with stochastic variational inference to deal with the increasing size of datasets with MapReduce.

However, extracting meaningful topics from a crawled document collection is a challenging task because the crawled documents are large in size and number. In order to solve the scalability problem, the open-source Hadoop platform with MapReduce framework is used to distribute the processing and to increase the computation of variational EM. This paper continues the work proposed in [12] and attempts to implement a scalable MapReduce CTM with variational EM algorithm to analyse the crawled full-text documents collection.

The main contributions of this paper are as follows:

- Implementing the variational EM algorithm for MapReduce CTM in a Hadoop cluster that is able to automatically discover the latent topics.
- Evaluating the results of proposed MapReduce CTM with another topic model and comparing the topic coherences of both models.

The remainder of this study has been structured as follows. In section 2, the theorical background of CTM is briefly explained. Next, section 3 presents the detailed workflow of the proposed approach. The experimental results on the crawled dataset are described in section 4. Then, section 5 discusses the conclusion and future works of this research.

## 2. THEORY BACKGROUND

The Correlated Topic Model (CTM) is a generative model to find the patterns of words in documents, to reveal the latent semantic themes of a collection of documents and to describe how these themes are distributed over individual texts [3]. CTM, one of the statistical topic models, is popular in natural language processing community to handle large amount of unstructured documents collection and is applied in many domains, such as images [8, 18], web services [10], computer vision [15] and text analysis [16, 5].

CTM assumes that the words of each document originate from a mixture over a set of latent topics. Again, each topic is modeled as a distribution over a set of words, i.e., the vocabulary. The key of CTM is the logistic normal distribution. CTM exhibits the correlations between latent topics through the covariance matrix of logistic normal distribution. As a consequence, the logistic normal adds complexity to the inferencing process of CTM. The understanding of CTM is that a document consists of many topics with different proportions and different topics have different distributions over the vocabulary. The graphical model representation of CTM is illustrated in Figure 1. The rectangles denote the replicated structure, and only the shaded node, the words of the documents, is observed.

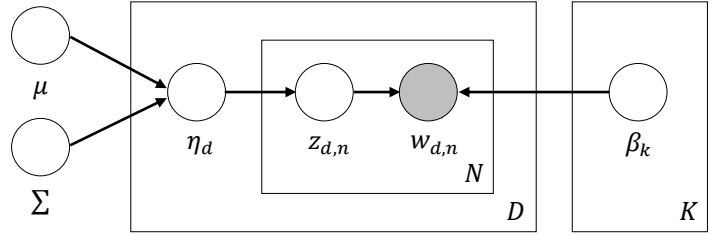

Figure 1. Graphical model of CTM

Given a collection of documents $D$, a $K$-dimensional Normal distribution of mean and covariance matrix $N(\mu, \Sigma)$ and some topics $K$, CTM assumes that the documents are generated according to the following generative process:

1. For each topic $k$, choose a distribution over the vocabulary $\beta_k \sim N(\mu, \Sigma)$.
2. For each document $d$,
    a. Choose a distribution over the topics $\eta_d \sim N(\mu, \Sigma)$.
    b. For each word $n$,
        i. Choose a topic assignment $z_{d,n}$ from $Mult(f(\eta_d))$.
        ii. Choose a word $w_{d,n}$ from $Mult(\beta_{z_{d,n}})$.

Topic proportion $\theta$ for each document is obtained from the logistic normal transformation,

$$\theta = f(\eta) = \frac{\exp\{\eta\}}{\sum_i \exp\{\eta_i\}} \qquad (1)$$

The word distribution per topic $\beta_k$ and topic distribution per document $\eta_d$ of CTM are difficult to compute directly, but various inference algorithms have been implemented to figure out this difficulty expeditiously. To learn the parameters of CTM, a variational Expectation-Maximization (EM) is used for inferencing [3]. The two procedures in variational EM consists of posterior inferencing of variational parameters and model parameters estimation.

Given an observed document $w$ and model parameters $\{\beta, \mu, \Sigma\}$, the posterior distribution of the latent variables $p(\eta, z \mid w, \beta, \mu, \Sigma)$ is intractable to compute. Jensen's inequality is used to bound the log probability of a document,

$$Log\, p(w_N \mid \mu, \Sigma, \beta) \geq E_q\left[log\, p(\eta \mid \mu, \Sigma)\right] + \sum_{n=1}^{N} E_q\left[log\, p(z_n \mid \eta)\right] +$$
$$\sum_{n=1}^{N} E_q\left[log\, p(w_n \mid z_n, \beta)\right] + H(q) \qquad (2)$$

where $H(q)$ denotes the entropy of variational distribution. For the posterior inferencing, the variational parameters are added to obtain the approximation of lower-bound on the likelihood of each document. Then, the variational distribution is set to,

$$q(\eta, z \mid \lambda, \nu^2, \phi) = \prod_{i=1}^{K} q(\eta_i \mid \lambda_i, \nu_i^2) \prod_{n=1}^{N} q(z_n \mid \phi_n) \tag{3}$$

where $(\lambda, \nu^2)$ is variational mean and covariance of normal distribution, $\phi$ is a variational multinomial distribution.

Given a collection of documents, the parameter estimation maximizes the likelihood of the whole documents collection by using a variational expectation-maximization (EM) algorithm. In the E-step, a variational inference for each document is performed to maximize the bound with respect to the variational parameters $\{\lambda, \nu^2$ and $\phi\}$. In the M-step, the bound is maximized with respect to the model parameters $\{\mu, \Sigma, \beta\}$. The E-step and M-step are executed repeatedly until the bound on the likelihood converges. The detailed explanations of posterior inference and parameter estimation can be found in [4].

## 3. PROPOSED METHODOLOGY

The proposed approach is composed of three phases: data gathering, data pre-processing and topic extraction via the MapReduce CTM. When the latent topics are discovered, the topic evaluation is performed using the UCI and UMass topic coherence measures. Figure 2 depicts the block diagram of proposed model implemented in Hadoop MapReduce framework.

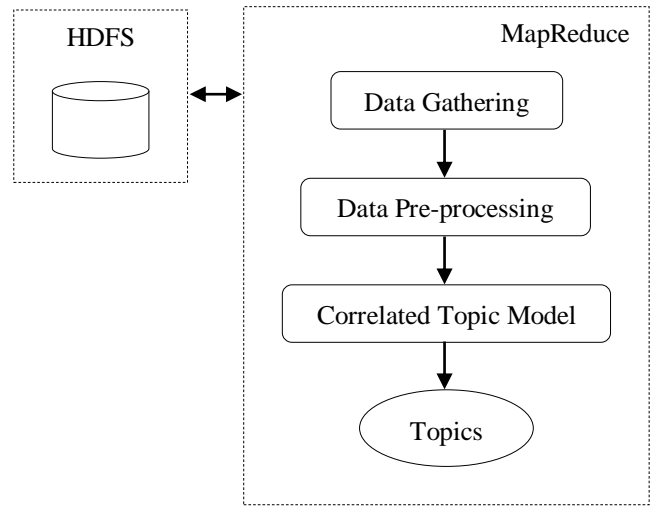

Figure 2. Block diagram of proposed approach

### 3.1. Data Gathering

The digital documents are gathered from the PLOS ONE digital library [19]. The PLOS ONE provides access to academic contents in any disciplines within science and medicine. A web crawler is developed in Java to read and gather the research documents ending in .pdf extension from the multidisciplinary library. The textual contents of each crawled document are extracted by applying the Apache PDFBox Java library. The extracted text data are uploaded and stored in Hadoop Distributed File System (HDFS) to perform further data pre-processing tasks and to learn the latent themes represented via the latent topics.

## 3.2. Data Pre-processing

The data pre-processing represents a critical role prior to topic extraction. For the three pre-processing tasks, the proposed approach uses the Map and Reduce functionalities. The input to the data pre-processing phase is the extracted text data stored in HDFS.

### 3.2.1. Words Extraction

From the raw text files, each line in each document are split into tokens. Then, the algorithm extracts only the words whose length are between 4 and 20 in Map function, and counts the occurrences of those words in Reduce function. The procedure of the words extraction process is described in Figure 3.

---

**Procedure: Words Extraction**

```
// key: key, value: document contents
method Map (LongWritable key, Text value) {
   for each word w in value
      w ← w.replaceAll("[^A-Za-z]+$", "").trim();
      if (w.length() < 4 || w.length() > 20)
         w ← w.replaceAll(w, "").replaceAll("\\s"+ " ").trim();
      endif
      Emit (w, one);
   endfor
}
// key: word, values: list of counts
method Reduce (Text key, Iterator values) {
   int result ← 0;
   for each value v in values
      result += v;
   endfor
   Emit (key, result);
}
```

---

Figure 3.  Procedure of words extraction

### 3.2.2. Stopwords Elimination

In this step, the words which occur less than five times in the dataset are removed. Stopwords which appear redundantly in almost every document and words without semantic meaning are also eliminated to speed up the topic extraction process. Figure 4 shows the procedure for stopwords elimination.

---

**Procedure: Stopwords Elimination**

```
Read stopword file from DistributedCache
// key: key, value: word, count
method Map (LongWritable key, Text value) {
   for each word w in value
      c ← extractInt(w);
      if (c >= 5 && !w.matches("[0-9]+") && !w.isEmpty())
         if (!stopWordList.contains(w))
            Emit (w, null);
         endif
      endif
   endfor
}
```

---

Figure 4.  Procedure of stopwords elimination

### 3.2.3. Spell Checking

After removing the stopwords, the spell checking is performed based on the dictionary file. The procedure of spell-checking process is summarized in Figure 5. For training CTM model, spell checking has a significant influence on the model's results.

---

**Procedure: Spell Checking**

---

```
Read dictionary file from DistributedCache
// key: key, value: word
method Map (LongWritable key, Text value) {
    for each word w in value
        if (!dictionaryList.contains(w))
            Emit (w, null);
        endif
    endfor
}
```

---

Figure 5. Procedure of spell checking

## 3.3. Correlated Topic Model in MapReduce Framework

Given a pre-processed text dataset, a MapReduce CTM is trained to learn the underlying themes that represent that corpus. The word-topic probability distributions and the topic-document probability distributions are computed in this phase. The variational inference of CTM in [4] is adopted to extract the topics from the full-text collection. In this work, the variational EM algorithm for CTM over MapReduce framework is implemented to handle the volume of documents collection. The topic representation of documents allows to summarize the whole documents collection without prior knowledge. In other words, it provides an interpretable latent structure of items so that to understand by humans.

The entire variational EM algorithm is divided into three parts: the Driver, the Mapper and the Reducer classes. The Driver class takes the control of the whole inference process and the responsibility of submitting the MapReduce job to the Hadoop cluster for execution. It first accepts the input dataset from HDFS and divides it into fixed-sized pieces called input splits. The Driver also takes the responsibility to initialize the model parameters $\{\mu, \Sigma, \beta_K\}$ and the variational parameters $\{\lambda_i, v_i^2, \zeta, \phi_{n,i}\}$.

The number of topics $K$ is user specified, and the corpus $D$ is determined by the data. For the variational EM iteration, the E-step is executed in the Mapper class and the M-step is executed in the Reducer class. The procedures for the Mapper and Reducer classes of MapReduce CTM are summarized in Figure 6 and Figure 7, respectively. A pair of Map and Reduce functions constitutes a single iteration of the variational EM algorithm. After each MapReduce iteration, the Driver updates $\beta$, $\mu$ and $\Sigma$.

---

**Procedure: Mapper Class**

---

//key: documentID, value: document contents
method Map (Intwritable key, Document value) {
repeat
  for $n = 1$ to $N_d$
    for $i = 1$ to $K$
      Update $\zeta$ with $\zeta = \sum_i exp(\lambda_i + v_i^2/2)$
      Update $\phi_{n,i}$ with $\phi_{n,i} = exp(\lambda_i)\,\beta_{i,n}$
    endfor
  endfor
  Update $\lambda_i$ with $dL/d\lambda = -\sum^{-1}(\lambda - \mu) + \sum_{n=1}^{N}\phi_{n,1:K} - (N/\zeta)\,exp(\lambda + v^2/2)$
  Update $v_i^2$ with $dL/dv_i^2 = -\sum_{ii}^{-1}/2 - (N/2\zeta)\,exp(\lambda + v_i^2/2) + 1/(2v_i^2)$
until convergence
Emit (key, likelihoods of variational parameters);
 }

---

Figure 6. Procedure of Mapper Class

For the Mapper class given in Figure 6, the MapReduce framework creates a new Map task for each input split. Since the input files are smaller than the HDFS split size, the number of mappers is equal to the number of input files. The Map function reads each record from the input dataset and maps input key-value pairs to intermediate key-value pairs. The objective of Mapper is to update and estimate the variational parameters for each document.

---

**Procedure: Reducer Class**

---

//key: key, values: list of values
method Reduce (key, Iterator values) {
  for $d = 1$ to $D$
    for $i = 1$ to $K$
      Update $\beta_i$ with $\beta_i = \sum_d \phi_{d,i}\, n_d$
    endfor
    Update $\mu$ with $\mu = \frac{1}{D}\sum_d \lambda_d$
    Update $\Sigma$ with $\Sigma = \frac{1}{D}\left(\sum_d diag(v_d^2) + \sum_d(\lambda_d - \mu)(\lambda_d - \mu)^T\right)$
  endfor
  Emit (key, model parameters);
 }

---

Figure 7. Procedure of Reducer Class

As in the Reducer class described in Figure 7, each Reduce task receives the intermediate output produced from the Map task and performs operation on the list of values against each key. The Reduce function emits the final output key-value pairs which are stored in HDFS. The objective of Reducer is to update the model parameters.

## 3.4. Topic Coherence Metrics

Since topics are not assured to be well interpretable to the coherence judgements of the humans, the topic coherence metrics are applied to reveal the semantic relatedness of the topics in order to measure the effectiveness of topic model. For the evaluations of the extracted topics from the MapReduce CTM, the two topic coherence measures [11], UCI and UMass, are used in this paper. The coherence of a single topic is scored by measuring the degree of semantic similarity between its high scoring words. Thus, the coherence of a topic model is computed by taking a mean of the coherence score per topic for all topics contained in the model.

The UCI coherence measure based on Pointwise Mutual Information (PMI) is described as follows:

$$C_{UCI} = \frac{2}{N \cdot (N-1)} \sum_{i=1}^{N-1} \sum_{j=i+1}^{N} PMI(w_i, w_j) \qquad (4)$$

where

$$PMI(w_i, w_j) = log \frac{P(w_i, w_j) + \varepsilon}{P(w_i) \cdot P(w_j)} \qquad (5)$$

where the probabilities are computed by counting the word co-occurrence. The UMass coherence measure is defined as:

$$C_{UMass} = \frac{2}{N \cdot (N-1)} \sum_{i=2}^{N} \sum_{j=1}^{i-1} log \frac{P(w_i, w_j) + \varepsilon}{P(w_j)} \qquad (6)$$

where the probabilities are derived by using the document co-occurrence counts. The smoothing parameter $\varepsilon$ is used to avoid taking the log of zero for the words that are never cooccurred.

## 4. EXPERIMENTAL RESULTS

### 4.1. Environmental Setup

The environmental setting was executed on a host computer running Microsoft Windows 10 OS. The experiments are run in a Hadoop cluster consisting of 3 nodes with 1 master and 2 slaves. All experiments are implemented using Java and Apache Hadoop 2.7.1 installed on Ubuntu 16.04 OS. The hardware profile of the host machine is a dual-core 2.70GHz CPU, 16GB of RAM and 1TB hard disk. The master node has 6GB of RAM and 150GB hard disk. For each slave node, 3GB of RAM and 100GB hard disk.

### 4.2. Dataset

The experiments are carried out based on the dataset crawled from PLOS ONE digital library with a time frame of 5 hours period. The dataset contains 148 full-text documents containing 407,309 total number of sentences and 2,696,316 total number of words. After the pre-processing of the dataset, the cleaned dataset is stored in HDFS which contains a number of 164,266 sentences and a total of 62,279 words. For the extraction of vocabulary, all stopwords and all infrequent and misspelling words are eliminated. The vocabulary is learned from the dataset and the size of the vocabulary is 3,729 words.

### 4.3. Topic Model Results

On the cleaned PLOS ONE dataset, the MapReduce CTM with variational EM algorithm is executed to extract the 10 topics. Table 1 presents the 10 topics extracted from the PLOS ONE dataset using MapReduce CTM. Each line is a topic composed of top 10 words semantically related with different degrees of relatedness. At the moment, the number of topics is arbitrarily set to 10 before investigating the optimal number of topics for the dataset.

Table 1.  Extracted 10 topics for PLOS ONE.

| Topics | Top 10 words |
|---|---|
| Topic 1 | privacy arctic density months effort planning range dose protection mobile |
| Topic 2 | dose mobile range confidentiality environment protection percent blood software method |
| Topic 3 | mobile future floodplain method economic protection percent integer syphilis node |
| Topic 4 | syphilis range method economic strategy environment months arctic raster plot |
| Topic 5 | future raster extent plot incident method description protection syphilis economic |
| Topic 6 | percent effort method item range incident protection description code economic |
| Topic 7 | range blood method protection effort code description plot economic arctic |
| Topic 8 | syphilis arctic training confidentiality future mobile percent detection acid range |
| Topic 9 | dose future plot percent economic confidentiality months range meeting description |
| Topic 10 | privacy syphilis legislation mobile department extent economic effort taxonomy method |

From the results in Table 1, it can be seen that, the words 'range', 'economic' and 'method' can be found in 7 topics and the word 'protection' in 6 topics, and so on. Many words are repeated in multiple topics showing that the number of topics set to 10 is too large for the PLOS ONE dataset. Therefore, it is important to identify the number of topics for the dataset when training a topic model.

In the next section, the evaluations of the topics are performed to identify the optimal number of topics because the CTM model itself cannot verify the optimal number of topics. Choosing the optimal number of topics depends on the nature of dataset. When too many topics are derived from the topic model, it may get over fitted which is not expected at all. On the other hand, extracting too few topics does not make sense too.

### 4.4. Topic Coherence Evaluations

To investigate the optimal number of topics discovered by the proposed MapReduce CTM, the two topic coherence measures, UCI and UMass, are used during the experiments. The proposed model is evaluated by changing the number of topics in order to select the optimal number of topics for the dataset.

For the experiments, with $\varepsilon$ set to 1.0E-12, the scores of topic coherences are significantly decreased towards the higher negative values. Then, setting $\varepsilon$ to 1.0E-6 gives the higher scores of UCI and UMass indicating that the generated topics have the better topic coherences.

Table 2.  Coherence scores of MapReduce CTM for PLOS ONE.

| Number of topics | UCI | UMass |
|---|---|---|
| 5 | **3.5993** | -3.994 |
| 6 | 1.8481 | -3.4385 |
| 7 | 1.6605 | -3.4759 |
| 8 | 2.3035 | -3.4861 |
| 9 | 1.6792 | -3.044 |
| 10 | 1.5913 | **-2.8284** |

Table 2 describes the UCI and UMass coherence measures of MapReduce CTM with different numbers of topics (from 5 to 10) for PLOS ONE dataset using the external Wikipedia dataset. From Table 2, it can be found that the scores of UCI and UMass measures are the highest at number of topics 5 and 10, respectively. For a topic model, a higher topic coherence score means that it contains more reasonable topics where each topic contains the most probable words that are frequently co-occur together. Hence, these numbers of topics are selected as the optimal number of topics for the PLOS ONE dataset because of having the highest UCI and UMass topic coherence scores.

Table 3.  Extracted 5 topics ordered by UCI scores of each topic for PLOS ONE.

| Topics | Top 10 words | UCI |
|--------|--------------|-----|
| Topic 3 | anchor, appendices, breast, supplemental, registry, temp, transaction, ozone, authority, gestation | 4.8769 |
| Topic 2 | signature, outlook, cent, breast, morbidity, reproductive, specification, procedure, shelf, protein | 3.4004 |
| Topic 5 | republic, overlap, frame, addendum, registry, oxford, spice, reproductive, veterinary, shipping | 3.3783 |
| Topic 1 | filename, reserved, directory, procedure, welfare, stem, discovery, reflect, origin, race | 3.2697 |
| Topic 4 | injection, shelf, peak, prospective, registry, organ, radii, authority, greenhouse, loop | 3.0714 |

Table 4.  Extracted 10 topics ordered by UMass scores of each topic for PLOS ONE.

| Topics | Top 10 words | UMass |
|--------|--------------|-------|
| Topic 6 | percent, effort, method, item, range, incident, protection, description, code, economic | -2.2101 |
| Topic 9 | dose, future, plot, percent, economic, confidentiality, months, range, meeting, description | -2.2323 |
| Topic 2 | dose, mobile, range, confidentiality, environment, protection, percent, blood, software, method | -2.4871 |
| Topic 10 | privacy, syphilis, legislation, mobile, department, extent, economic, effort, taxonomy, method | -2.6528 |
| Topic 7 | range, blood, method, protection, effort, code, description, plot, economic, arctic | -2.7864 |
| Topic 1 | privacy, arctic, density, months, effort, planning, range, dose, protection, mobile | -2.8052 |
| Topic 4 | syphilis, range, method, economic, strategy, environment, months, arctic, raster, plot | -2.8139 |
| Topic 8 | syphilis, arctic, training, confidentiality, future, mobile, percent, detection, acid, range | -3.0173 |
| Topic 5 | future, raster, extent, plot, incident, method, description, protection, syphilis, economic | -3.1697 |
| Topic 3 | mobile, future, floodplain, method, economic, protection, percent, integer, syphilis, node | -4.1094 |

Table 3 and Table 4 present the extracted topics for PLOS ONE dataset ordered by UCI and UMass scores, respectively. From the results in Table 4, the words 'range', 'economic' and 'method' can be found in 7 topics and the word 'protection' in 6 topics, and so on. Many words are repeated in multiple topics showing that the number of topics set to 10 is too large for the PLOS ONE dataset. However, the topics in Table 3 cover the terms relating to the aspects of 'file system in a computer', 'environmental authority' and 'structure of organism'. Therefore, after manually evaluating the topics, the number of optimal topics for PLOS ONE dataset is chosen as 5 topics due to the more understandable results with small number of topics used.

Table 5. Coherence scores of Mr. LDA for PLOS ONE.

| Number of topics | UCI | UMass |
|---|---|---|
| 5 | 0.7462 | -2.5143 |
| 6 | 0.9781 | -2.6019 |
| 7 | 0.7727 | -2.1707 |
| 8 | 0.9649 | -2.6888 |
| 9 | 0.9418 | -2.4868 |
| 10 | 0.7957 | -2.3252 |

The performance of MapReduce CTM is compared with Mr. LDA for the PLOS ONE dataset. The Mr. LDA [9] is a distributed large-scale topic modeling algorithm using variational inference technique and is implemented in MapReduce framework. Table 5 describes the UCI and UMass coherence measures of different numbers of topics (varying from 5 to 10) computed by the Mr. LDA for the PLOS ONE dataset.

Figure 8 illustrates the UCI and UMass coherence measures computed by MapReduce CTM and Mr. LDA for the PLOS ONE dataset.

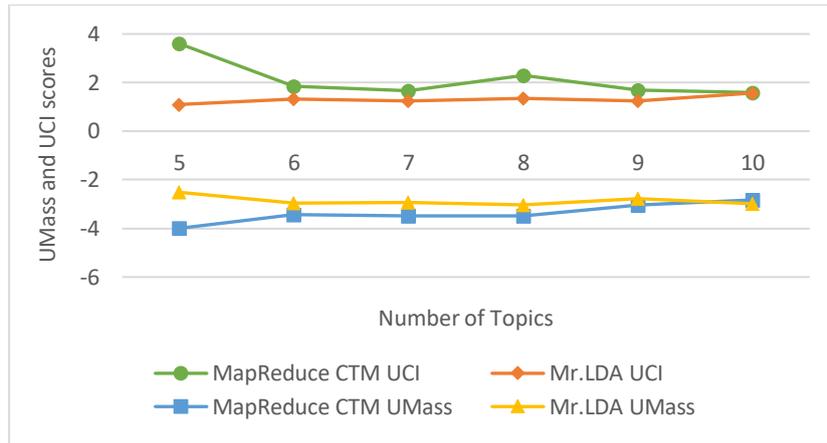

Figure 8. Coherence scores of Mr. LDA and MapReduce CTM

For the UCI scores, the scores of MapReduce CTM does not increase significantly than Mr. LDA except at the number of topics 5. At this point, the UCI score reached its peak for MapReduce CTM model, that is, the model produces more reasonable topics containing more semantically related words than Mr. LDA. For the number of topics 6, 7, 8 and 9, the UCI scores are slightly higher than Mr. LDA.

For the UMass scores, MapReduce CTM has slightly lower UMass scores than Mr. LDA except at topics 5. This is because of the reason that more redundant topic words are generated at number of topics 5. The UMass score of MapReduce CTM at topics 10 is the highest among other number of topics and has a very little rise than Mr. LDA.

## 4.5. Training Time Comparison

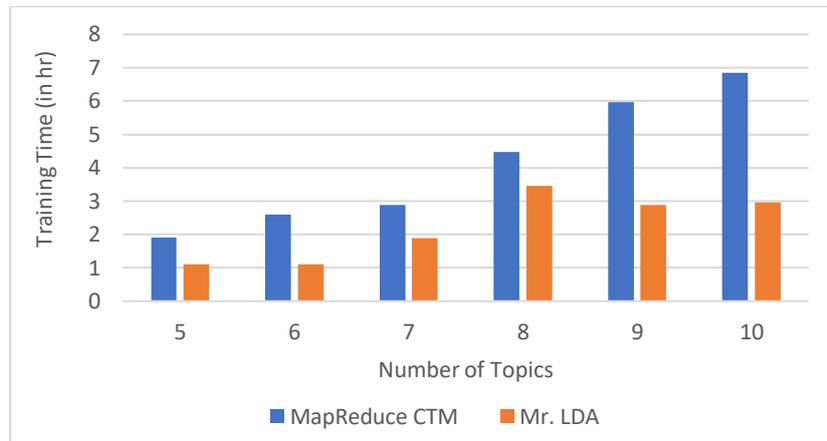

Figure 9. Training time of MapReduce CTM and Mr. LDA

Figure 9 shows the comparison between the varying number of topics and the time taken for training of the PLOS ONE dataset using MapReduce CTM and Mr. LDA topic models. It was observed that the training time increases with the increase in the number of topics for the two topic models. Moreover, the training time of MapReduce CTM is significantly higher than that of Mr. LDA because MapReduce CTM contains more parameters and requires more computations for the correlations of topics.

## 5. Conclusion

In this paper, the MapReduce CTM with variational EM algorithm is implemented for the crawled documents collection in a Hadoop cluster to extract the latent topics in order to understand the whole documents collection. For the experiments, the full-text documents are crawled from the PLOS ONE digital library to increase the quality of extracted topics. The performance of the proposed MapReduce CTM model is evaluated in terms of UCI and UMass coherence measures. According to the topic coherence evaluations, although the proposed MapReduce CTM does not have relatively better performance when extracting topics for a particular dataset, it has a comparable performance as a topic modeling method. The results show that the topic coherences of MapReduce CTM model slightly perform better than Mr. LDA in most of the cases measured with UCI score and performs marginally worse than Mr. LDA in some cases measured with UMass score.

This work mainly focuses on the extraction of latent topics from the crawled documents collection. There are still many further works that are needed to be done. In the future, the work will be emphasized on increasing the size of documents collection and improving the performance of variational EM algorithm. Furthermore, the MapReduce CTM will be developed to be applicable for improved information extraction.